\documentclass[twocolumn,english,aps, prl, showpacs]{revtex4}
\usepackage{graphicx}
\usepackage{graphics}
\usepackage{times}
\usepackage{amstext}
\usepackage{bbm}
\usepackage{babel}

\begin{document}

\title{Scalable quantum computation via local control of only two qubits}

\author{Daniel Burgarth$^{1,2}$}

\author{Koji Maruyama$^{2}$ }

\author{Michael Murphy$^{3}$ }

\author{Simone Montangero$^{3}$}

\author{Tommaso Calarco$^{3,4}$ }

\author{Franco Nori$^{2,5}$ }

\author{Martin B. Plenio$^{1}$}

\affiliation{$^{1}$IMS and QOLS, Imperial College, London SW7 2PG, UK }

\affiliation{$^{2}$Advanced Science Institute, The Institute of Physical and
Chemical Research (RIKEN), Wako-shi, Saitama 351-0198, Japan}

\affiliation{$^{3}$Institut für Quanteninformationsverarbeitung, Universität
Ulm, D-89069 Ulm, Germany}

\affiliation{$^{4}$ECT{*}, 38050 Villazzano (TN), Italy}

\affiliation{$^{5}$Physics Department, University of Michigan, Ann Arbor, Michigan,
48109, USA}
\begin{abstract}
We apply quantum control techniques to control a large
spin chain by only acting on two qubits at one of its ends, thereby
implementing universal quantum computation by a combination of quantum
gates on the latter and swap operations across the chain. It is shown
that the control sequences can be computed and implemented efficiently.
We discuss the application of these ideas to physical systems such
as superconducting qubits in which full control of long chains is
challenging.
\end{abstract}


\maketitle
Controlling quantum systems at will has been an aspiration for physicists
for a long time. Achieving quantum control not only clears the path
towards a thorough understanding of quantum mechanics, but it also
allows the exploration of novel devices whose functions are based
on exotic quantum mechanical effects. Among others, the future success
of quantum information processing depends largely on our ability to
tame many-body quantum systems that are highly fragile. Although the
progress of technology allows us to manipulate a small number of quanta
quite well, controlling larger systems still represents a considerable
challenge. Unless we overcome difficulties towards the control over
large many-body systems, the benefits we can enjoy with the `quantumness'
will be severely limited. 

One encouraging result is that almost any coupled quantum system
is controllable in principle, even by steering a single particle
only \cite{Lloyd2004}. The question of controllability in this type
of situation can be described by the theory of quantum control \cite{D'Alessandro2008},
which uses Lie algebraic arguments. This is interesting from the theoretical
point of view; however, can it be practically useful from the quantum
computing perspective? Problems we need to contemplate before attempting
to build a large quantum computer using quantum control are as follows.
First, the control criterion is generally not computable for large
systems. Second, even if the question of controllability can be answered
positively for specific systems \cite{Burgarth2008}, the precise
sequence of actual controls (or `control pulses') are generally not
computable. And third, even if they can be computed, the theory of
control tells us nothing about the overall \emph{duration} of the
control pulses to achieve a given task, and it might take far too
long to be practically relevant.

The usual approach to circumvent these problems focuses on systems
that are sufficiently small, so that we do not already require a quantum
computer to check their controllability and to design control pulses.
In such a case, the theory of time optimal control \cite{CONTR}
can be used to achieve impressive improvements in terms of total time or type of pulses required in comparison with 
the standard gate model. More complicated desired operations on larger
systems are then decomposed (`compiled') into sequences of smaller
ones. Yet, the feasibility of this approach is ultimately limited
by the power of our classical computers, therefore constrained to
low-dimensional many-body systems only.

The first step to achieve control over large-scale systems by controlling
only a few particles was done in the context of quantum state transfer
in spin chains. In these simple models,
the issues mentioned above were avoided by restricting the analysis to the
subspace with a single excitation. Most of the proposed
schemes for quantum state transfer \cite{QST,UQI}
are not actually based on the framework of control theory, but on
smart tricks from various fields of physics, often using classical
intuition about the dynamics. Also, the theory of optimal control
was recently applied to state transfer \cite{CONTRQST, CONTRQST1}.%
\begin{figure}
\includegraphics[width=1\columnwidth]{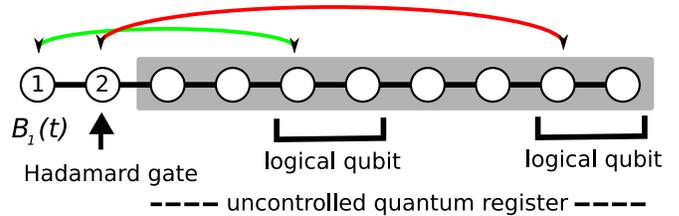}\caption{\label{fig:Our-scheme-for}(color online) Our approach for universal quantum computation
works on a chain of $N$ spins. By modulating
the magnetic field $B_{1}(t)$ on qubit $1$, we induce information
transfer and swap gates on the chain (red and green lines). The states of the qubits
from the uncontrolled register can be brought to the controlled part. There, the
gates from a quantum algorithm are performed by local operations.
Afterward, the (modified) states are swapped back into their original
position.}

\end{figure}

A few of these proposals were then applied to more general tasks than
state transfer; that is, how to use spin chains for entanglement purification
\cite{Maruyama2008}, cloning transformations \cite{Chiara2005},
or even for fully fledged quantum computation with little control
\cite{UQI}. Thus, such schemes are no longer restricted
to a small subspace of the full Hilbert space. These schemes for quantum
computation use clever methods to design the control pulses analytically,
i.e., without relying on control theory, but at the price of limitations
on their applicability. In particular, they assume specific coupling
parameters and design of the system. Nonetheless, such methods are
intriguing from a theoretical perspective (e.g., in relation to complexity
questions) and give hope for the feasibility of quantum control of
larger systems.

The goal of this paper is to efficiently compute control pulses
for a large system, using the full Hilbert space, and to show that
the duration of the pulses scales efficiently (i.e., polynomially)
with the system size. There are various trade-offs to consider. For
instance, if we allow more means for external control and/or higher
ability in designing the system Hamiltonian, the problem will become
more tractable theoretically (on paper), but more demanding and less
relevant from an experimental perspective. 

Here, we present a solution to achieve feasible control both
theoretically and experimentally. We will use a Hamiltonian that can
be efficiently diagonalized for large systems through the Jordan-Wigner
transformation. The control pulses are applied only to the \emph{first
two} spins of a chain (see Fig.~\ref{fig:Our-scheme-for}). The control
consists of two parts: one where we will use the Jordan-Wigner to
efficiently compute and control the information transfer through the
chain (thus using it as a quantum data bus), and a second one where
we will use some local gates acting on the chain end to implement two-qubit
operations. To be efficiently computable, these local gates need to
be fast with respect to the natural dynamics of the chain. Combining
the two actions allows us to implement any unitary operation described
in the gate model. The interplay of the Jordan-Wigner picture for
the information transfer and the gate model in the canonical basis
will also require some tricks to avoid the accumulation of (uncomputable)
phases.

\paragraph*{System and information transfer problem.---}

We consider a chain of $N$ spin-$1/2$ particles coupled by the Hamiltonian

\[H=\mbox{$\textstyle \frac{1}{2}$}\sum_{n=1}^{N-1}c_{n}[(1+\gamma)XX+(1-\gamma)YY]_{n,n+1}+\sum_{n=1}^NB_{n}Z_{n},\]
where $X,Y,Z$ are the Pauli matrices, the $c_{n}$ are generic coupling constants and $B_{n}$ represent
a magnetic field. 
 Varying the parameter $\gamma$
allows to encompass a wide range of experimentally relevant Hamiltonians
\cite{SOLID,Muller2008,Garcia-Ripoll2003}, including
the transverse Ising model ($\gamma=1$; for this case we require the fields $B_n\neq0$) and the $XX$ model ($\gamma=0$).
We assume that the value of 
the first magnetic field, $B_{1},$ can be controlled externally.
This control will
be used to induce information transfer on the chain and realize  swap
gates %
\footnote{We remark that a swap gate 
between two sites $k$ and $l$ is much more specific than achieving
quantum state transfer \cite{QST} between them. A swap performs
\emph{bi-directional} quantum state transfer and does not change the
state of the remaining sites. %
} between arbitrary spins and the two `control' spins $1,2$ at one
chain end. Hence such swap gates are steered \emph{indirectly} by
only acting on the first qubit. The tunable parameter will also
be used for applying unitaries on the control spins.

In order to focus on the main idea we now present our method for $\gamma=0$
and $B_{n}=0$ for $n>1$.  The general case follows
along the same lines, though being more technically involved. Our
first task is to show that by only tuning $B_{1}(t),$ we can perform
swap gates between arbitrary pairs of qubits. First we rewrite the
Hamiltonian, using the Jordan-Wigner transformation $a_{n}=\sigma_{n}^{+}\prod_{m<n}Z_{m}$,
into $H=\sum_{n=1}^{N-1}c_{n}\{a_{n}^{\dag}a_{n+1}+a_{n+1}^{\dag}a_{n}\}.$
The operators $a_{n}$ obey the canonical commutation relations $\{a_{n},a_{m}^{\dag}\}=\delta_{nm}$
and $\{a_{n},a_{m}\}=0.$ The term we control by modulating $B_{1}(t)$
is $h_1=Z_{1}=1-2a_{1}^{\dag}a_{1}.$
From the theory of quantum control \cite{D'Alessandro2008} we know
that the reachable set of unitary time-evolution operators on the
chain can be obtained from computing the \emph{dynamical Lie algebra}
generated by $ih_1$ and $iH.$ It contains all possible commutators
of these operators, of any order, and their real linear combinations.
For example, it contains the anti-Hermitian operators $ih_{12}\equiv =[ih_1,[ih_1,iH]]/(4c_{1})=i(a_{1}^{\dag}a_{2}+a_{2}^{\dag}a_{1}),$ $ih_{13}\equiv\left[iH,ih_{12}\right]/c_{2}=a_{1}^{\dag}a_{3}-a_{3}^{\dag}a_{1}$
and $ih_{23}\equiv\left[ih_{12},ih_{13}\right]=i(a_{2}^{\dag}a_{3}+a_{3}^{\dag}a_{2}).$
We observe that taking the commutator with $h_{12}$ exchanges the
index $1$ of $h_{13}$ with $2.$ Taking the commutator with $iH$
we find that $ih_{14}\equiv\left[ih_{13},iH\right]+ic_{1}h_{23}-ic_{2}h_{12}=i(a_{1}^{\dag}a_{4}+a_{4}^{\dag}a_{1})$
and $ih_{24}\equiv a_{2}^{\dag}a_{4}-a_{4}^{\dag}a_{2}$ are also
elements of the dynamical Lie algebra. Hence the effect of taking
the commutator with $H$ is raising the index of the $h_{kl}.$ Generalizing
this, we find that the algebra contains the elements
$ih_{kl},$ with $k<l,$ $ih_{kl}\equiv a_{k}^{\dag}a_{l}-a_{l}^{\dag}a_{k}$
for $(k-l)$ even, $ih_{kl}\equiv i(a_{k}^{\dag}a_{l}+a_{l}^{\dag}a_{k})$
for $(k-l)$ odd, and  $h_k=Z_{k}=1-2a_{k}^{\dag}a_{k}.$
From control theory \cite{D'Alessandro2008} we
thus know that the time evolution operators $\exp(-3\pi ih_{kl}/2)$
(which will turn out to be very similar to swap gates) can be achieved
through tuning $B_{1}(t).$ The main point is that because both $h_{1}$
and $H$ are free-Fermion Hamiltonians, the corresponding control
functions can be computed \emph{efficiently} in a $2N$-dimensional
space (we will do so explicitly later).
Ultimately, we need to transform the operators back to the canonical
spin representation. Using $a_{k}^{\dag}a_{l}=-\sigma_{k}^{-}\sigma_{l}^{+}\prod_{k<j<l}\sigma_{j}^{z},$
we find $\exp(-\pi ih_{kl}/2)=(|00\rangle_{kl}\langle00|+|11\rangle_{kl}\langle 11|)\otimes \openone +(|01\rangle_{kl}\langle10|-|10\rangle_{kl}\langle01|)\otimes L_{kl}$ for $(k-l)$ odd. The operator $L_{kl}=\prod_{k<j<l}\sigma_{j}^{z}$
arises from the non-local tail of the Jordan-Wigner transformation
and act only on the state of the spins \emph{between} $k$ and $l.$
They are \emph{controlled} by the state of the qubits $k,j$ being in the
odd parity sector. 

In order to use the chain as a quantum data bus, our goal is to
implement\emph{ swap gates}
$S_{kl}=|00\rangle_{kl}\langle00|+|11\rangle_{kl}\langle11|+|10\rangle_{kl}\langle01|+|01\rangle_{kl}\langle10|$,
so the fact that we have achieved some modified operators with different
phases on $k,l$ instead, and also the controlled non-local phases
$L_{kl},$ could potentially be worrisome. It is worth pointing out that
the $XX$ interaction is not capable of generating swap gates of the
(physical) qubits.  We will use a method suggested in \cite{ALASTAIR}
that allows us to tackle these complications.  That is, rather than
using the physical qubits, we encode in \emph{logical} qubits,
consisting of two neighbouring physical qubits each. They are encoded in
the odd parity subspace $|01\rangle, |10\rangle$.  Swapping a logical
qubit $n$ to the control end of the chain then consists of two physical
swaps $\exp(-\pi ih_{1 \, 2n-1}/2)$ and $\exp(-\pi ih_{2 \,
  2n}/2)$. Since both physical swaps give the same phases, the
resulting operation is indeed a full logical swap. Any single-qubit
operation on the logical qubits can be implemented by bringing the
target qubit to the control end, performing the gate there, and bringing
it back again. If we modulate the magnetic field $B_1(t)$, we could
equally decide to perform single logical qubit gates directly, without bringing them to the control end. This is possible
because $\exp(-ih_{2n-1 \, 2n}t)$ in the physical picture translates to
$\exp{(-i X_{L,n}t )}$ in the logical picture, and because $Z_{2n-1}$ is in
the algebra generated by $Z_1$, which allows us to perform the operation $\exp{(-i
  Z_{2n-1}t)} =\exp{(-i Z_{L,n}t)}$.

For quantum computation, we need to be able to perform at least one entangling two-qubit operation. We choose a controlled-Z operation, as this can be performed by operating only on half of two logical qubits. That is, to perform a controlled-Z between logical qubit $n$ and $m$, we bring the physical qubits $2n-1$ and $2m-1$ to the control end, perform a controlled-Z between them, and bring them back. It is easy to check that again all unwanted phases cancel out. The controlled-Z  could
not be efficiently computed in the interplay with the many-body Hamiltonian
$H$, because it cannot be generated by a quadratic Hamiltonian in the Jordan-Wigner picture. Therefore, this gate has to be implemented on
a time-scale $t_{g}$ much faster than the natural evolution of the
chain, i.e., $t_{g}\ll\min_{j}\{1/c_{j}\}.$ We can soften this requirement by using control theory to generate $\exp{(-i Z_1 X_2 t)}$ by modulating $\beta_1(t)Y_1$ (this is a linear term in the Jordan-Wigner picture), and then using a fast Hadamard gate on the second site to obtain  $\exp{(-i Z_1 Z_2 t)}$, which, together with $\exp{(-i Z_1 t)}$ and $\exp{(-iZ_2 t)}$, gives the controlled-Z gate. This gives us quite a remarkable conclusion: except for a fast Hadamard on a single qubit, all other controls required for quantum computation can be computed efficiently in the Jordan-Wigner picture.

\paragraph*{\label{sec:Efficiency}Efficiency.---}

The crucial question left open in the above is: how long does it
actually take to implement the gates? Unfortunately, the theory of
quantum control does not provide a general answer, though some
interesting progress was recently reported \cite{D'Alessandro2009}. In
order to evaluate the efficiency, we have numerically simulated a range
of chain lengths and studied the scaling of the logical swap operation
time $T$ with the (physical) chain length $N$. For the control, we
choose the local magnetic field $B_1(t)$ at the first site.
We set the coupling strength constant,
namely $c_{n}=J\ \forall\ n$.  To provide evidence of a polynomial
scaling, we set the simulation time $T_{N}=(N-1)^{2},$ (all times are
given in units of $1/J$ and $\hbar=1$) and verify for each $N$ that we can find a
specific $B_{1}^{\ast} (t)$ that performs the logical swap %
\footnote{Note that the time $T_N$ is the time it takes to perform a
  physical swap, the logical swap will take twice this time.}.
\begin{figure}
\begin{centering}
\includegraphics[width=1\columnwidth]{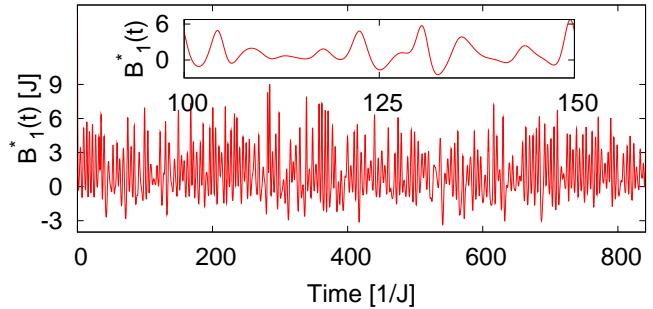} 
\par\end{centering}

\caption{(color online) The optimized function $B_{1}^{\ast}(t)$ that
  produces the physical swap between $n = 1$ and $n = 29$ for the
  chain $N=30$ in a time $T_{30}$ (in units of $1/J$). The inset
  shows a zoom to confirm that the time-scale of the oscillations of
  $B_{1}^{\ast}(t)$ are slower than the dynamics of the chain.}

\centering{}\label{fig:pulse} 
\end{figure}
 We quantify our success by calculating the error of the operation
$\varepsilon=1-F$, where $F=(|\mathrm{tr}{U^{\dagger}U_{g}}|/N)^{2}$
is the fidelity between the time evolution $U$ and the goal unitary
$U_{g}$. We find the function $B_{1}(t)$ using techniques from optimal
control theory \cite{Khaneja2001,CONTR,D'Alessandro2008}.
Briefly, the procedure is as follows: (1) an initial guess is made
for the function $B_{1}(t)$; (2) we run the optimal control algorithm
to generate a new $B_{1}(t)$ which decreases the error of our operation;
(3) steps 1 and 2 are iterated until the final error reaches a defined
threshold $\varepsilon$. In practice, it suffices to choose a threshold which
is of the same order of magnitude as the error introduced by 
the Hadamard gate. 

\begin{figure}
\begin{centering}
\includegraphics[width=1\columnwidth]{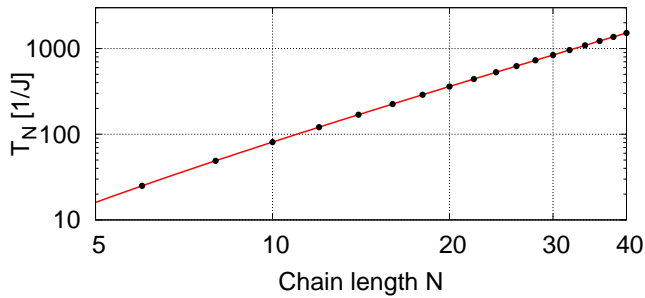} 
\par\end{centering}

\caption{(color online) The swap operation time $T_{N}$ versus the chain length
$N$: the red continuous line is $(N-1)^{2}$ while the black dots are
the lengths for which we numerically verified this scaling with an
error $\varepsilon < 10^{-4}$. }

\centering{}\label{fig:scale} 
\end{figure}

If the algorithm converges for each $N$ and the corresponding $T_N$,
giving the optimal pulse sequence $B_{1}^{\ast}(t)$, then we can assert
that the scaling of the operation time is at least as good as
$T_N=(N-1)^{2}$, up to a given precision. We initialize the algorithm
with a guess for the function $B_{1}(t)$, e.g., $B_{1}(t)=1$. This is,
of course, a rather poor choice for performing the swap gate. However,
the optimization algorithm exponentially improves the choice of
$B_{1}(t)$ reaching, after several thousand iterations, the desired
precision.  A typical optimized function $B_{1}^{\ast}(t)$ is given in
Figure~\ref{fig:pulse}.  It is worth pointing out that although the
pulse \emph{looks} like it is oscillating very rapidly, its oscillations
are slower than the natural dynamics of the chain, thus the control
function does not need to be fast. Indeed, a Fourier analysis confirmed
that only frequencies of up to $\sim0.02J$ are required. Furthermore, such
pulses are robust against small fluctuations \cite{CONTR}. The results of
the optimizations for different chain lengths are shown in
Figure~\ref{fig:scale}: the desired quadratic scaling has been clearly
achieved.  We stress here that the chosen scaling law $T_{N}$ may not
necessarily describe the shortest time on which the physical swap gate
can be performed.  However, the dynamical Lie algebra of quasi-free
fermions has a dimension of the order $N^2$, indicating that such
scaling might be optimal.  Note that even though here we focused on
swapping between the chain ends, which one expects to takes the longest
time, we checked that general swaps between any pair of logical qubits
can be achieved in (at least) the same time-scale (data not shown).

A final remark on the robustness against imperfections of the results
presented is needed, as perfect homogeneity or even fine tuning of
individual couplings might be very hard to achieve in most systems and
might prevent information transfer schemes to work \cite{ANDERS}.  In
the previous analysis, for simplicity, we considered a chain with uniform
couplings $c_{n}=J$ for $n>1$, but the results for arbitrary couplings
are similar as long as localization effects can be neglected.  We assume that the disordered Hamiltonian
parameters are known, because they can be estimated efficiently by controlling only the chain end \cite{Burgarth2009,Franco2009}. We
checked that the optimization with off-site disorder of $10\%$,
uniformly distributed, leads to the same results,for most realizations, until at least
$N=40$.

\paragraph*{Conclusion.---}

We have shown how to \emph{efficiently} compute control pulses for
large spin chains described by a vast class of Hamiltonians. The pulses
are computed for an $2N$-dimensional system but can be applied to
the full $2^{N}$-dimensional system. Full quantum computation is
possible by only controlling \emph{two} spins at one end of the chain.
The only price for this indirect control is that the quantum computation
takes quadratically longer than for direct control. Given the large
benefit of requiring so little control for a quantum computer, we
believe that this scheme would be very useful for future implementations.
As a further application, we remark that our proposal can also be applied
to use the spin chain as a quantum memory, storing qubits by
moving states from the controlled part to the rest of the chain and
by applying control pulses on qubits $1,2$ to achieve the identity
on the register (effectively switching off the chain Hamiltonian).
For future studies, we would additionally like to probe the ultimate
limit at which one may perform swaps using optimal control 
\cite{Khaneja2001,CONTR,D'Alessandro2008, CONTRQST1},
and try to obtain simple (possibly analytic \cite{D'Alessandro2009})
pulses.

As a possible application of our proposal, we note that in Josephson
qubit implementations, much progress has been reported on the control
of two-qubit gates \cite{SOLID}, but that controlling
and reading out many qubits is very hard. In fact, it is expected
to be difficult to construct fully addressable long arrays of Josephson
junctions while it is conceivable to produce long chains of qubits
with always-on interaction where only one or two qubits are fully
controllable and readable %
\footnote{Hans Mooij, private communication%
}. In those systems the decoherence time is (optimistically) $\sim1000$
times larger than the time-scale of the inter-qubit coupling \cite{Kakuyanagi2007,Grajcar2006},
which would make our scheme applicable for up to $\sim30$ qubits
to achieve a single swap gate (though, of course, much less to achieve
full computation). Hence, the optimal control ideas presented here have
the potential to address a serious limitation in such implementations
and thus open a novel avenue towards quantum information processing
in solid state devices.

We thank V. Giovannetti, S. Schirmer and P. Pemberton-Ross for fruitful discussions and A. Kay for useful feedback. We acknowledge support by the EPSRC grant
EP/F043678/1 (DB), Incentive Research Grant of RIKEN (KM),
US NSA,  LPS, ARO, NSF Grant No. EIA-0130383 (FN),
EU under the Integrated Project SCALA (TC) and Contract No. MRTN-CT-2006-035369
(EMALI) (MM),  DFG SFB TRR21 (SM and TC),
 Royal Society Wolfson Research Merit Award, Integrated Project
QAP and the EU STREP
project HIP (MP).

\end{document}